\documentclass[runningheads]{llncs}

\newif\ifanonymous
\anonymousfalse
\usepackage{float,siunitx} 
\usepackage{amsmath,amssymb,amsfonts}
\usepackage{booktabs,caption,subcaption}
\usepackage{algorithmicx,setspace}
\usepackage[noend]{algpseudocode}
\usepackage{graphics,graphicx,xcolor}
\usepackage[utf8]{inputenc}
\usepackage{authblk}

\usepackage{pst-node}% http://ctan.org/pkg/pst-node
\usepackage{multido}% http://ctan.org/pkg/multido

\usepackage{varioref,hyperref}
\hypersetup{
  linkcolor  = red,
  citecolor  = green,
  urlcolor   = blue,
  colorlinks = true,
}
\usepackage[nameinlink,capitalize]{cleveref}

\usepackage{tikz}
\usetikzlibrary{positioning,chains,fit,shapes,calc}
\usepackage[most]{tcolorbox}
\usepackage{comment}
\usepackage[section]{placeins}

\usepackage{mathtools}

\newcommand{\glabel}[1]{\{X^b\}}

\def\bitcoinA{%
  \leavevmode
  \vtop{\offinterlineskip %\bfseries
    \setbox0=\hbox{B}%
    \setbox2=\hbox to\wd0{\hfil\hskip-.03em
    \vrule height .3ex width .15ex\hskip .08em
    \vrule height .3ex width .15ex\hfil}
    \vbox{\copy2\box0}\box2}}
\usepackage[utf8]{inputenc}
\usepackage{hyperref}

\hypersetup{
    colorlinks=true,
    linkcolor=blue,
    filecolor=magenta,      
    urlcolor=blue,
    citecolor=blue,
    pdftitle={Short Paper: The Spatiotemporal Scaling Laws of Bitcoin Transactions} 
}

\ifanonymous
\title{Short Paper: The Spatiotemporal Scaling Laws of Bitcoin Transactions}
\else
\title{The Spatiotemporal Scaling Laws of \\Bitcoin Transactions}
\fi
\titlerunning{The Spatiotemporal Scaling Laws of Bitcoin Transactions}
\ifanonymous
\author{}
\institute{}
\else
\author{Lajos Kelemen\inst{1}\and István András Seres\inst{1}\and Ágnes Backhausz\inst{1,2}}
\institute{Eötvös Loránd University, Budapest, Hungary\and Rényi Institute of Mathematics, Budapest, Hungary}
\authorrunning{L. Kelemen et al.}
\fi
\date{\today}

\begin{document}

\maketitle
\begin{abstract}
    This study, to the best of our knowledge for the first time, delves into the spatiotemporal dynamics of Bitcoin transactions, shedding light on the scaling laws governing its geographic usage. Leveraging a dataset of IP addresses and Bitcoin addresses spanning from October 2013 to December 2013, we explore the geospatial patterns unique to Bitcoin. Motivated by the needs of cryptocurrency businesses, regulatory clarity, and network science inquiries, we make several contributions. Firstly, we empirically characterize Bitcoin transactions' spatiotemporal scaling laws, providing insights into its spending behaviours. Secondly, we introduce a Markovian model that effectively approximates Bitcoin's observed spatiotemporal patterns, revealing economic connections among user groups in the Bitcoin ecosystem. Our measurements and model shed light on the inhomogeneous structure of the network: although Bitcoin is designed to be decentralized, there are significant geographical differences in the distribution of user activity, which has consequences for all participants and possible (regulatory) control over the system.

\end{abstract}

\section{Where is George? And Satoshi?}
How do people spend their money? Where and when do they send transactions? What are the scaling laws of the spatiotemporal spending patterns of users in major financial systems? Is there any significant difference in the spending patterns of cash and digital currencies? How much time elapses between two consecutive transactions of a user? How many kilometres do paper bills travel during their lifetime? Given the difficulty of obtaining this type of geospatial data from payment processors, these questions seem unanswerable at scale. 

Fortunately, while obtaining such geospatial data is challenging, we have found compelling answers in the case of cash. In December 1998, an influential website~\url{https://www.wheresgeorge.com/} was launched as a semi-serious game to track the movements of US dollar bills. Players were invited to enter the serial number of the bills, the time, and the place they received bills with this specific stamp on them. Over the years, a large database emerged that tracked the movement of hundreds of thousands of paper bills across the United States.

In 2006, network scientists Brockmann, Hufnagel, and Geisel conducted a comprehensive analysis of this database~\cite{brockmann2006scaling}. They found that the trajectories of paper bills roughly follow a two-dimensional random walk known as Lévy flight. 
In particular, the distances between subsequent observations of paper bills follow a power law distribution. Intuitively, bills predominantly jump small distances (e.g., remain in a city). At the same time, with non-negligible probability, they observed lengthy jumps (e.g., the bill's owner travelled across states). Similarly, the waiting times between transactions also follow a power law distribution. 

Now, turning our attention to digital currencies, particularly Bitcoin~\cite{nakamoto2008bitcoin}, we raise questions about whether similar scaling laws apply and how they might differ. One would expect significant differences from the distributions observed in physical cash. Most importantly, digital currencies do not have physical limitations, i.e., one can easily send electronic transactions across continents. It would be surprising if the spatiotemporal spending patterns of digital currencies would also follow a two-dimensional Lévy flight. But then, what kind of distribution do they follow? Would it be possible to assess this?

Traditional centralized financial institutions, e.g., banks and credit card companies, extensively collect, analyze, and apply real-time geospatial data of their customers. Geographic information systems manage resources and optimize bank branch networks and marketing efforts. While centralized financial systems heavily rely on geospatial data for various purposes, obtaining and analyzing this data for academic research poses significant challenges due to its sensitivity and the reluctance of financial institutions to share it. This is partly due to the risks for re-identification using credit card metadata~\cite{de2015unique}.

On the other hand, decentralized digital currencies, such as Bitcoin, publicly offer this type of user data. However, it is not easy to map IP addresses to cryptocurrency addresses as it requires a large-scale measurement on the peer-to-peer (P2P) network of Bitcoin. There are multiple known vulnerabilities on the Bitcoin P2P network protocol (most of them are now patched) that could have allowed anyone to link IP addresses to Bitcoin transactions~\cite{apostolaki2021perimeter,biryukov2014deanonymisation,juhasz2018bayesian,koshy2014analysis}. These papers develop techniques to reduce the anonymity guarantees of the P2P network of Bitcoin. Still, they do not analyze the resulting data sets they obtained. Often, they run their deanonymization attacks solely on testnets due to ethical concerns~\cite{gao2021practical}. Fortunately, we obtained a substantial dataset consisting of $1797$ IP addresses and $20680$ Bitcoin addresses, dating back to 2013, courtesy of the authors of~\cite{juhasz2018bayesian}. This dataset is invaluable for our research.

\paragraph{Motivation for the study.} The following applications and network scientific questions motivate a deeper understanding of Bitcoin's geospatial scaling laws.
\begin{description}
    \item[Bussinesses Accepting Cryptos] Crypto companies, to grow and scale, need to know their product's usage better. In particular, they could have aggregate geospatial data about their customers. Such information allows companies to decide in which countries they should enable cryptocurrency payment options to reach more users or where to establish new cryptocurrency ATMs. 
    \item[Regulatory clarity] The wider cryptocurrency community can only hope for regulatory clarity if regulators understand Bitcoin's geospatial usage. Regulators might want to focus on parts of their corresponding country with high cryptocurrency activity for consumer protection and taxation purposes. 
    \item[Network scientific understanding and privacy] Finally, a high-quality geospatial database facilitates a deep network scientific understanding of Bitcoin and cryptocurrencies. We can assess Bitcoin's economic network's scaling laws, its (de)centralization, robustness, and privacy (mixing) characteristics.
\end{description}

\paragraph{Our contributions.} In this work, we provide the following contributions.
\begin{description}
\item[Bitcoin transactions' spatiotemporal scaling laws] We characterize Bitcoin transactions' spatiotemporal scaling laws in~\Cref{sec:bitcoinscalinglaws} using the aforementioned database. To the best of our knowledge, this is the first work to assess the spatiotemporal spending patterns of any cryptocurrency empirically.
\item[Markovian model of Bitcoin transactions] Bitcoin's spatiotemporal patterns do not lend themselves to be characterized as a simple two-dimensional random walk, e.g., Lévy flight. However, in~\Cref{sec:markovModel}, we can approximate well the observed spatiotemporal patterns of Bitcoin with a simple Markovian model that sheds light on the economic connections between various groups of users (i.e., miners, merchants, and users) in the Bitcoin ecosystem.
  \item[Open-source code] We applied a database from~\cite{juhasz2018bayesian} that contains spatiotemporal data on Bitcoin users from 2013. We publish our code at the following link:~\url{https://anonymous.4open.science/r/Scaling-Laws-of-Bitcoin-C834}. Due to privacy and ethical concerns of the applied data, we can provide access to the geospatial data upon request for reproduction and future research. 
\end{description}

The rest of this paper is organized as follows.  
In~\Cref{sec:bitcoinscalinglaws}, we introduce the database from~\cite{juhasz2018bayesian} used to measure the scaling laws of Bitcoin and present the results of our measurements. In~\Cref{sec:markovModel}, we describe a Markovian model of Bitcoin users' spatial sending patterns that explains the observed spatial behaviours. We conclude our work in~\Cref{sec:future} with future directions.
\section{The Spatiotemporal Patterns of Bitcoin Transactions}\label{sec:bitcoinscalinglaws}
In this Section, we describe the obtained data from Juhász et al.~\cite{juhasz2018bayesian} and thoroughly analyze their database through the lens of network science.

\subsection{The Collected Data}\label{sec:collecteddata}
The dataset we analyze in this study was collected by Juhász et al.~\cite{juhasz2018bayesian} during a comprehensive data collection campaign conducted between October 6th, 2013, and December 25th, 2013. During this two-month period, they operated $140$ Bitcoin nodes strategically distributed across various geographical regions. The campaign yielded an impressive corpus of data, recording approximately 300 million broadcast and relay events, encompassing $4,155,387$ transactions and identifying $124,498$ unique IP addresses. To ensure the reliability of the data, Juhász et al. employed a Bayesian approach to ascertain the actual originators of Bitcoin transactions, allowing them to establish meaningful connections between IP addresses and Bitcoin transactions. It is worth noting that only mappings with a probability of correctness exceeding $95\%$ were retained for our analysis. Our database consists of $101,342$ Bitcoin transactions, wherein both the sender's and receiver's IP addresses are known. While this may represent a small fraction ($\approx 2.44\%$) of the overall $\sim 4.15$ million Bitcoin transactions during the study period, it is statistically significant for our research. In contrast, we presume that Brockmann et al.~\cite{brockmann2006scaling} had access to a significantly smaller fraction of the total number of US dollar cash transactions. We acknowledge the potential use of network anonymity tools, such as Tor or i2p, by some users to obscure their actual IP addresses. However, it is important to note that these tools were not as prevalent during Bitcoin's early days, adding a layer of robustness to our dataset. This dataset serves as the foundation for our investigation, offering valuable insights into Bitcoin's spatiotemporal scaling laws and providing a unique window into the behaviour of cryptocurrency users during this critical period. %\agnes{esetleg valahova ebbe a bekezdesbe lehetne valami ilyesmit irni: As we will see, the users in our database cover a wide range regarding the number of transactions or the amount of Bitcoin spent, hence in spite of the smaller amount of data compared to the whole system and possible bias due to the unknown IP addresses, we can hope that we have a good sample of the network.}

\subsection{Spatiotemporal Patterns of Bitcoin Transactions}\label{sec:spatiotemporalBTC}
\begin{figure}[ht!]
\centering
\begin{subfigure}{.5\textwidth}
  \centering
  \includegraphics[width=\linewidth]{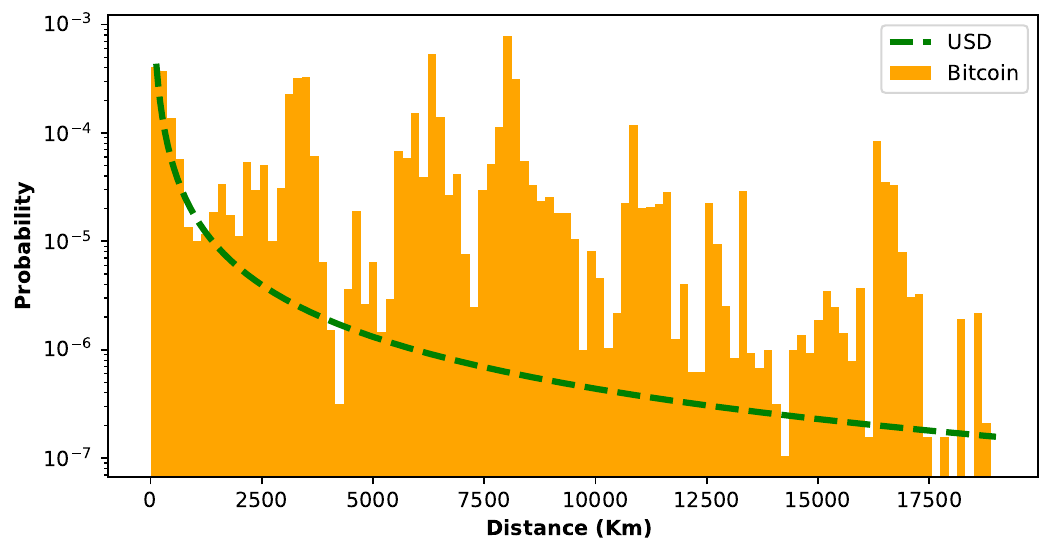}
\end{subfigure}%
\begin{subfigure}{.5\textwidth}
  \centering
  \includegraphics[width=\linewidth]{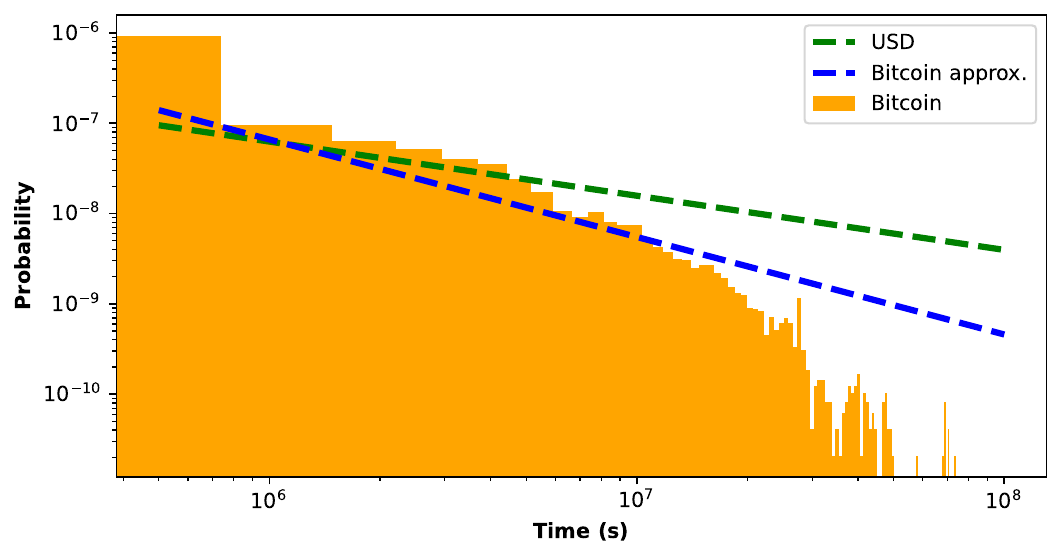}
\end{subfigure}
\caption{Distance and waiting time distributions of Bitcoin and US dollar as measured in~\cite{brockmann2006scaling}. Note the left figure is log-lin, while the right figure is log-log.}
\label{fig:distwaitingtimedists}
\end{figure}

Unlike physical cash, Bitcoin transactions' distance distribution does not follow a power-law distribution. Observe the much stronger tails of Bitcoin's transaction distance distribution  in~\Cref{fig:distwaitingtimedists} as opposed to the US dollar's power law distribution ($\sim x^{-1.59}$). The average distance a Bitcoin transaction covers is $5588.71$km with a median of $6236.6$km. In~\Cref{sec:markovModel}, we create a Markovian transaction model that explains and generates the same empirical distance distribution as observed in~\Cref{fig:distwaitingtimedists}. 

%%------------- BEGIN Dist and Waiting time distributions FIGURES -------------------------------
%%------------- END Dist and Waiting time distributions FIGURES -------------------------------
The elapsed time between the creation of a Bitcoin unspent transaction output (UTXO) and its spending as an input of a transaction is called~\emph{waiting time}. We found that just like in the case of US dollar bills ($\sim x^{-0.6}$), Bitcoin's waiting time distribution also follows a power law distribution ($\sim x^{-1.08}$), see~\Cref{fig:distwaitingtimedists}. The waiting time distribution has a mean of $18.44$ days and a median of $1.42$ days. In accordance with previous work~\cite{lischke2016analyzing}, we attribute the outstanding number of small waiting times to gambling activity, e.g., Satoshi Dice.

%%------------- BEGIN Time and Value time heatmaps FIGURES -------------------------------
\begin{figure}[ht!]
\centering
\begin{subfigure}{.5\textwidth}
  \centering
  \includegraphics[width=\linewidth]{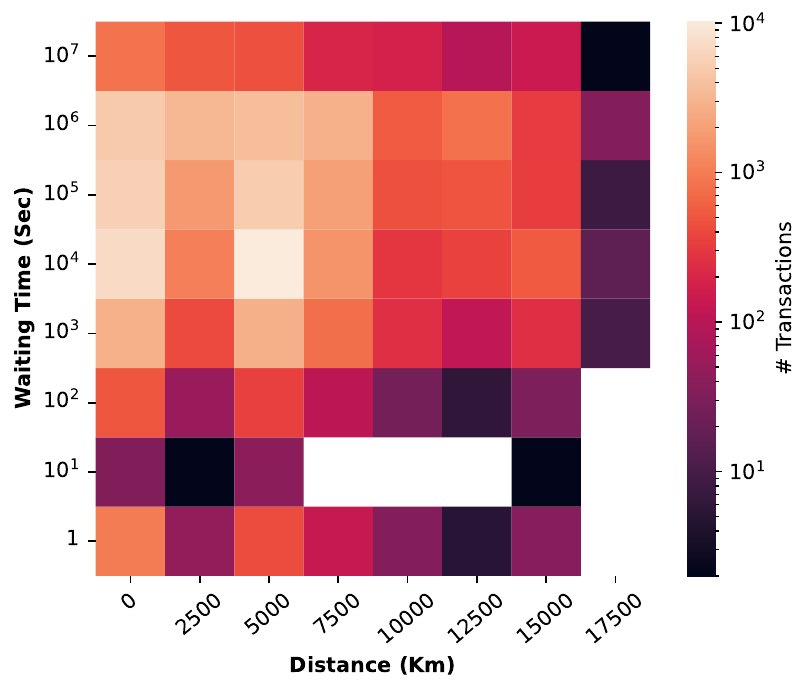}
\end{subfigure}%
\begin{subfigure}{.5\textwidth}
  \centering
  \includegraphics[width=\linewidth]{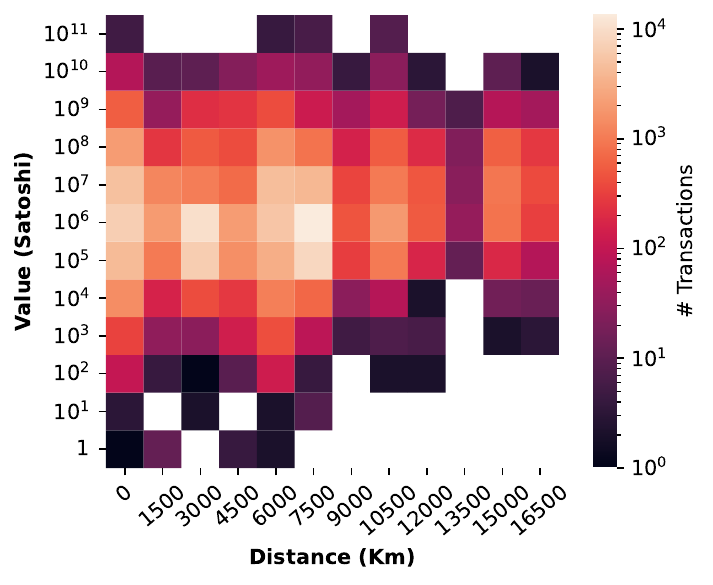}
\end{subfigure}
\caption{Transactions' waiting time (left) and value distribution (right) as a function of distance. Note both figures are log-lin, and the colour bar is also log. 
%Recall that the transactions' values and distances are independent, i.e., Pearson correlation $0.003$, Kendall's $\tau=0.021$, Spearman's $\rho=0.029$.
}
\label{fig:heatmaps}
\end{figure}
%%------------- END Time and Value heatmaps FIGURES -------------------------------

\paragraph{Independence of transaction value and distance.} We calculated the correlation of the transactions' transferred Bitcoin value and many spatial properties of the transactions, e.g., the distance between sender and receiver or the latitude and longitude of the sender. Importantly, but perhaps unsurprisingly, we found that the value of the transactions is uncorrelated with the distance between sender and receiver (correlation $0.003$). This is somewhat expected as Bitcoin transactions do not have physical limitations. Interestingly, the more southern the transaction's sender is, the longer the distance between the transaction's sender and receiver will be (correlation $-0.363$). This is because the countries in the global south transact regularly with the northern countries, e.g., Argentina with Germany. Similarly, we found that the distance between a transaction's sender and receiver and the waiting times are independent (correlation $0.03$).

%%------------- BEGIN Correlation FIGURES -------------------------------
%\begin{figure}
%\centering
%\begin{subfigure}{.5\textwidth}
  %\centering
  %\includegraphics[width=\linewidth]%{Figures/correlation.png}
%\end{subfigure}%
%\begin{subfigure}{.5\textwidth}
  %\centering
  %\includegraphics[width=\linewidth]{Figures/cikk_distval.pdf}
%\end{subfigure}
%\caption{Correlation between various spatial properties and the value of the Bitcoin transactions (left). The distribution of high-value transactions as a function of the distance between the transactions' sender and receiver (right).}
%\label{fig:correlations}
%\end{figure}
%%------------- END Correlation FIGURES -------------------------------

%%------------- BEGIN Btc World Map FIGURES -------------------------------
%\begin{figure}
 %   \centering
  %  \includegraphics[width = 10cm]{Figures/world_map.pdf}
   % \caption{A long chain of deanonymized transactions between various Bitcoin exchanges. Observe the long-distance jumps of the transactions.}
   % \label{wm}
%\end{figure}
%%------------- END Btc World Map FIGURES -------------------------------

%%------------- BEGIN Btc World Map Activity Heatmap FIGURES -------------------------------
%\begin{figure}
 %   \centering
  %  \includegraphics[width = 9cm]{Figures/heatmap_new.png}
   % \caption{Bitcoin transaction activity heatmap on the globe during the studied period, i.e., from October 2013 till December 2013.}
    %\label{fig:activity_heatmap}
%\end{figure}
%%------------- END Btc World Map Activity Heatmap FIGURES -------------------------------
\paragraph{Transaction activity and the Bitcoin user graph}
In the analyzed timeframe, Bitcoin's most active users were concentrated in regions such as the US East Coast, Europe, and Southeast Asia, with notable activity in China (see~\Cref{fig:markov_original2}). We further examined the Bitcoin user graph, identifying users by their Bitcoin addresses and tracking their transactions. Notably, the distribution of incoming and outgoing transactions from these users follows a power-law distribution (approximately $\sim x^{-1.51}$ and $\sim x^{-1.59}$, respectively). This finding aligns with earlier research~\cite{kondor2014rich} and suggests a concentration of influence among a subset of users – a phenomenon known as the ``Matthew effect". This observation raises questions about the decentralized nature often associated with cryptocurrencies.

\begin{figure}
\centering
\begin{subfigure}{.6\textwidth}
  \centering
  \includegraphics[width=\linewidth]{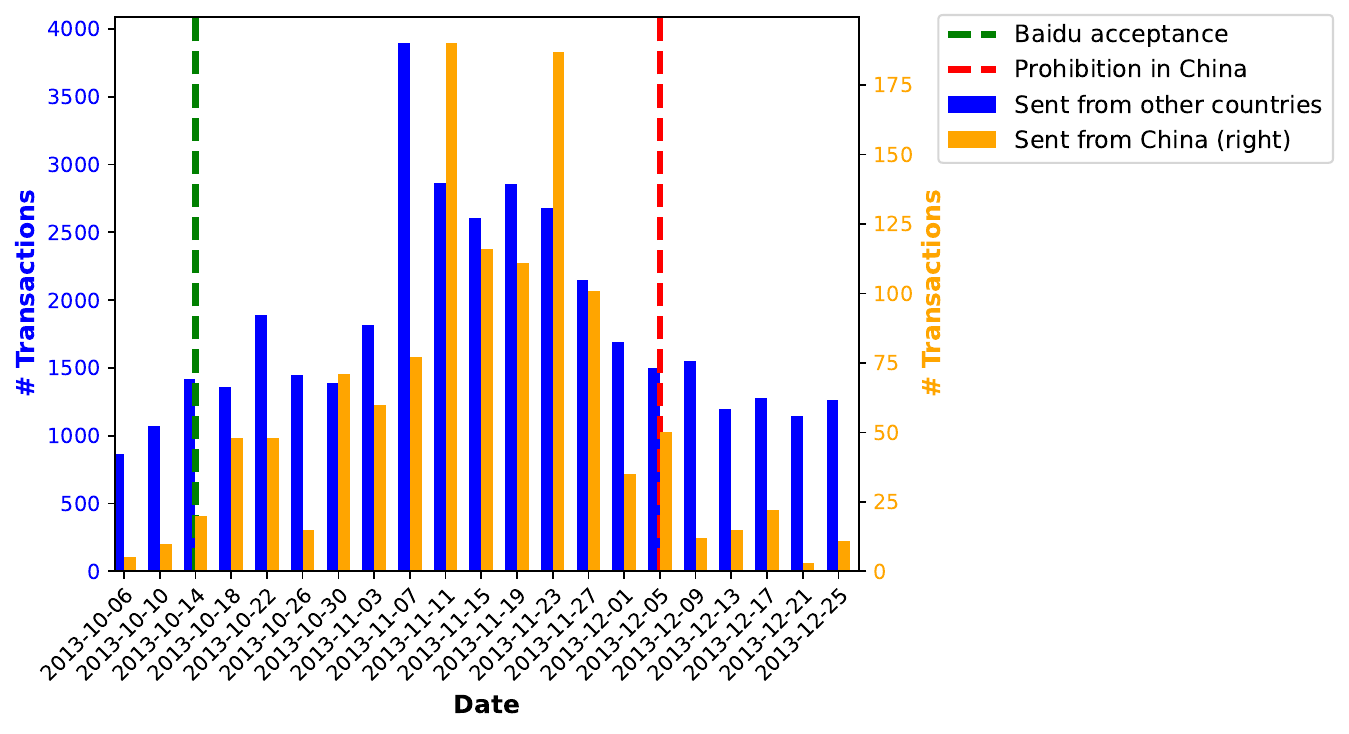}
\end{subfigure}%
\begin{subfigure}{.4\textwidth}
  \centering
  \includegraphics[width=\linewidth]{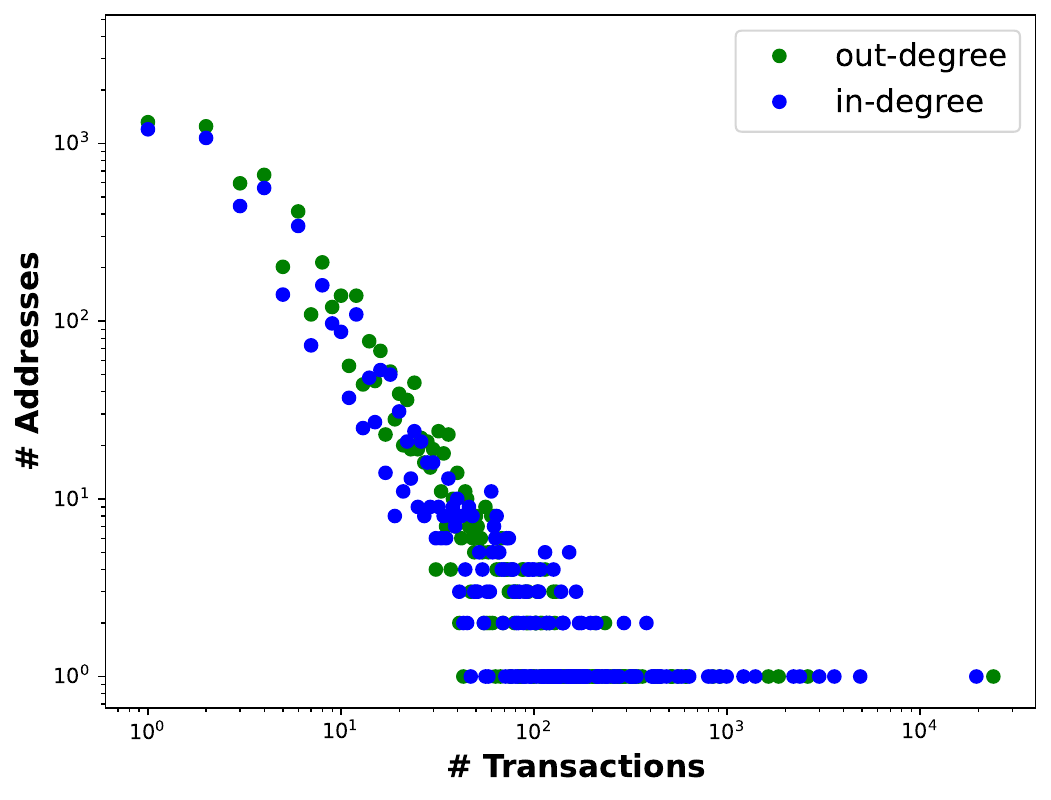}
\end{subfigure}
\caption{The number of sent Bitcoin transactions from the world and China (left). The in- and out-degree ($\sim x^{-1.51}$ and $\sim x^{-1.59}$) of the Bitcoin user graph (right).}
\label{fig:china_activity}
\end{figure}

\paragraph{Regulation and Bitcoin activity}
In October 2013, a significant development occurred when Baidu, China's largest search service, publicly announced its acceptance of Bitcoin as a payment method for its firewall and DDoS protection service~\cite{kapur13china}. Following this announcement, a notable surge in Bitcoin activity within China was observed, see~\Cref{fig:china_activity}. However, this surge was short-lived, as on December 5, 2013, the Chinese Communist Party declared Bitcoin to be an illegal currency in China and prohibited Chinese Bitcoin exchanges from accepting further renminbi deposits~\cite{kelion13bitcoin}. This regulatory action immediately and dramatically impacted Bitcoin transactions originating from Chinese IP addresses, as reflected in the data, see~\Cref{fig:china_activity}. These real-world events shaped the trajectory of Bitcoin activity in China and serve as a compelling illustration of the intricate relationship between regulatory decisions and cryptocurrency usage patterns.

%%------------- BEGIN China Activity -------------------------------
%%------------- END China Activity FIGURES -------------------------------
\section{A Markovian Model of Bitcoin Transactions}\label{sec:markovModel}
We observed in~\Cref{sec:spatiotemporalBTC} that the distribution of distances covered by Bitcoin transactions does not follow a simply characterizable two-dimensional (random) walk. To explain the empirical geospatial distribution of Bitcoin transactions, in this Section, we introduce a Markovian transaction model that generates the same spatial distribution of Bitcoin transactions as in~\Cref{fig:distwaitingtimedists} (left).

First, we group each Bitcoin address into one of the following three groups based on their activity in the observed period. The defining parameters of our Markovian model are established later as the result of an optimization problem.
\begin{description}
    \item[Miners] A node is classified as a miner if they had sent at least one transaction with value $\mathit{val}$ and participated in not more than $\mathsf{tx}_{miner}$ Bitcoin transactions either as a sender or a receiver.
    \item[Merchants/Service providers] Participated as sender or receiver in at least $\mathsf{tx}_{merch}$ transactions in the studied period. Examples include custodial Bitcoin exchanges, non-custodial wallets, or online casinos.
    \item[Users] If none of the above holds for a particular Bitcoin address in our database, then the Bitcoin network participant is deemed a ``regular'' user.
\end{description}
Furthermore, each participant is characterized by their geographic location according to their IP address. Specifically, users are assigned to the continent where they are based, i.e., America (AM), Asia (AS), and Europe (EU). In our database, there was no activity from Africa and only a few transactions from Australia and Oceania were added to the Asian user group. This resulted in nine user categories depending on the network participants' location (AM, AS, and EU) and assigned user type (miner, merchant, user). 

%%------------- BEGIN Approximate Distance Distribution FIGURES -------------------------------
\begin{figure}
    \centering
    \begin{subfigure}{.5\textwidth}
  \centering
    \includegraphics[width=\linewidth]{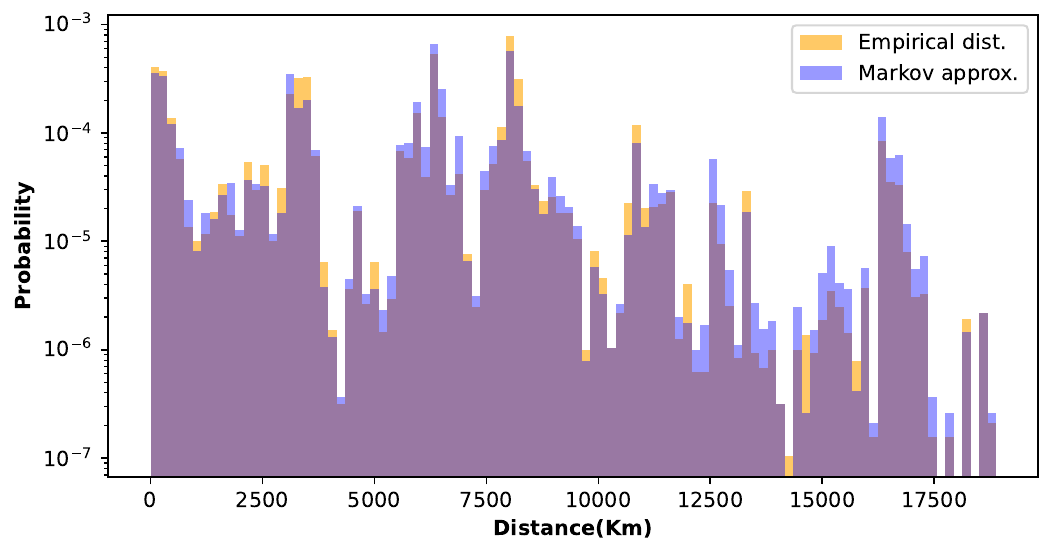}
    \end{subfigure}
    \begin{subfigure}{.48\textwidth}
        \includegraphics[width =\linewidth]{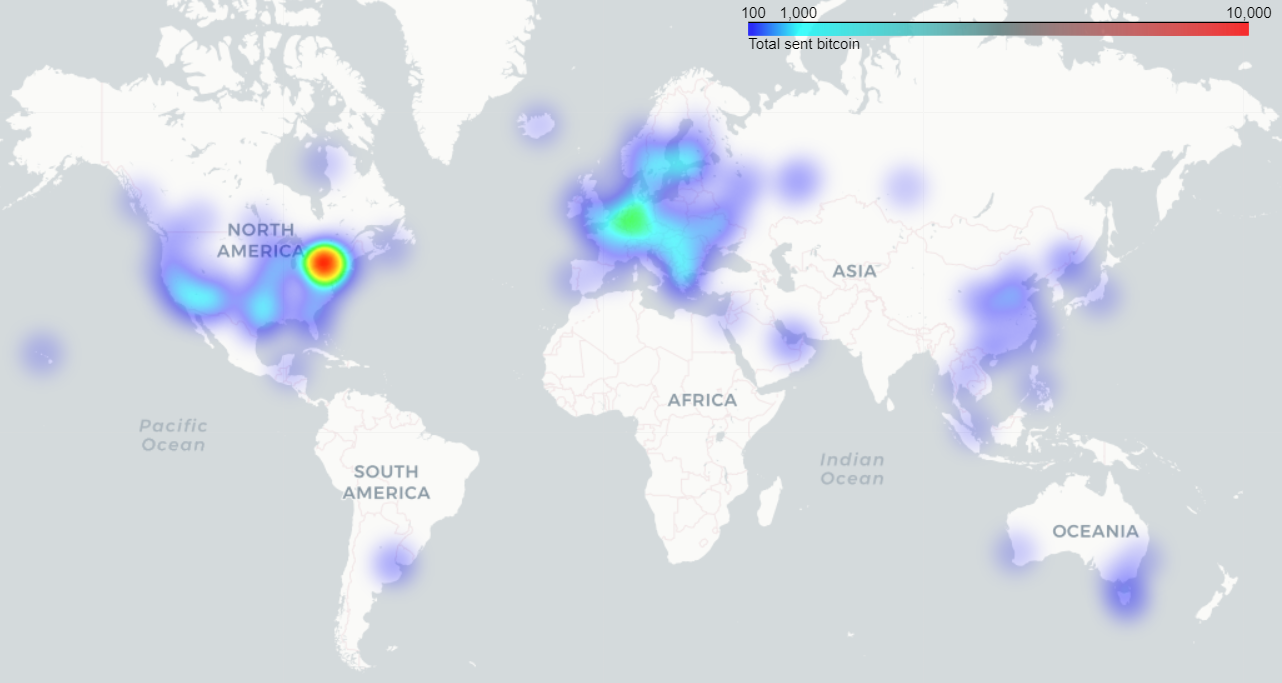}
    \end{subfigure}
    \caption{The empirical distance distribution (cf.~\Cref{fig:distwaitingtimedists} (left)) and the Markovian-generated for the optimal $\mathsf{params}=((15\bitcoinA,50,120))$  yielding the largest p-value $p=0.081$ in the Kolmogorov-Smirnov statistics between the two distributions (left). Bitcoin transaction activity heatmap during the studied period, i.e., from October to December 2013 (right), note similarity with~\cite{lischke2016analyzing}.}
    \label{fig:markov_original2}
\end{figure}
%%------------- END Approximate Distance Distribution FIGURES -------------------------------

We iterated over the parameter space $\mathsf{params}:=(\mathit{val},\mathsf{tx}_{miner},\mathsf{tx}_{merch})$ to find the optimal parametrization of our Markovian model, i.e.,  which $\mathsf{params}$ triplet minimizes the distance between the generated distance distributions and the empirical. We assigned network participants to the nine above-mentioned user groups for a fixed $\mathsf{params}$ triplet. The transition matrix for a given $\mathsf{params}$ list was defined from the relative transaction frequencies between the user groups that were assigned to the nine user groups. We generated $500$ transactions for each $\mathsf{params}$ triplet. For each $500$ transactions, we assigned a distance defined by the particular transition matrix. For each possible pair of a sender and a receiver from the nine groups, we determined the empirical distribution of the transactions' distance from the first group to the second one. Once we know the sender's and receiver's user groups for each transaction, we randomize the distance from this probability distribution. 
We computed the Kolmogorov-Smirnov (K-S) statistics~\cite{smirnov1948table} between the empirical distance distributions, see~\Cref{fig:markov_original2}, and the Markovian-generated. The largest p-value $p=0.081$ in the K-S statistics can be found for $\mathsf{params}=(15\bitcoinA,50,120)$. To validate the accuracy of our model, we examined its assignment of some well-known Bitcoin addresses. Notably, it correctly identified addresses associated with prominent entities like SatoshiDice and OkCoin, reinforcing its effectiveness in categorizing users.

%%------------- BEGIN Markov Transition Matrix FIGURES -------------------------------
\begin{figure}
\centering
\begin{subfigure}{.47\textwidth}
  \centering
  \includegraphics[width=\linewidth]{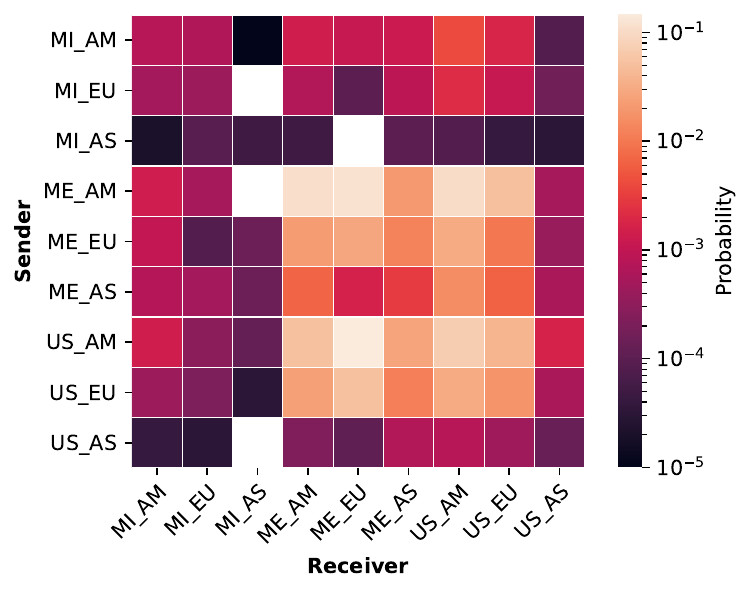}
\end{subfigure}%
\begin{subfigure}{.53\textwidth}
  \centering
  \includegraphics[width=\linewidth]{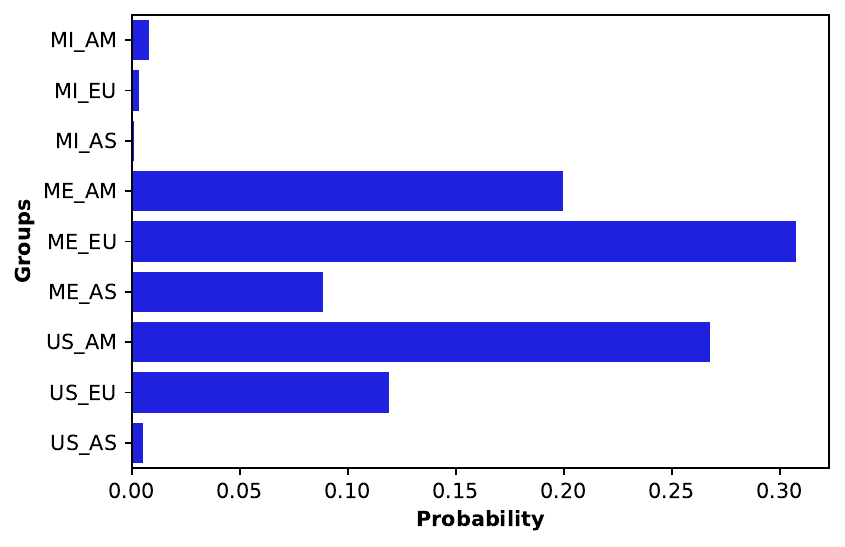}
\end{subfigure}
\caption{The relative frequency matrix with logarithmic probabilities (left) 
and stationary probabilities (right). Users are grouped into three classes: miners (MI), merchants (ME), and regular users (US). The three main prevalent geographic locations are denoted as America (AM), Asia (AS), and Europe (EU).}
\label{fig:markov_transitions}
\end{figure}
%%------------- END Markov Transition Matrix FIGURES -------------------------------
%%miért nem lehet egy eloszlással leírni, Amerika folytonos, itt bizonyos távolságok nem jelenhetnek meg, fontosak az interkontinentális tx-ek
We conclude that it seems elusive to describe Bitcoin's spatiotemporal scaling laws (e.g., the distance distribution of~\Cref{fig:distwaitingtimedists}) with a continuous probability distribution as it was possible for the US dollar~\cite{brockmann2006scaling}. We attribute this difficulty to the discontinuity of the observed distance distribution, e.g., certain distances cannot appear due to physical limitations, in particular, the special structure of the continents. Instead, we built a Markovian model that uses the intrinsic properties of the underlying network to generate the desired distance distribution.

We observe in~\Cref{fig:markov_transitions} that the relative frequencies differ significantly across the identified user groups, for example, high merchant and moderate miner activity. 
Although our model is rather simple, based on this, one can identify users with significantly different roles in the Bitcoin ecosystem: some are involved in many transactions, others are more like regular users. Hence, some kind of inhomogeneity appears despite the originally decentralized system of Bitcoin transactions. Similarly, we observe inhomogeneity across user groups in the stationary probabilities. In conclusion, our Markovian model provides valuable insights into Bitcoin transaction distributions, revealing notable inhomogeneities across user groups. These findings hold significant implications, particularly in the realm of regulations, as they highlight the need to focus regulatory efforts on user groups with the most substantial impact within the Bitcoin ecosystem.

%\lali{szerintem valahova még bele lehetne írni, hogy a deanonimizált címek a megfelelő kategóriába estek a csoportosítás során}

%\lali{a stacionárius eloszlásról lehetne még írni, hogy jól kihozza, hogy a minereknél nem marad a btc, és hogy EU meg Amerika a meghatozó szereplők, ahogyan más cikkek is leírják ezt}
\section{Conclusion and Future Directions}\label{sec:future}
In this study, we have delved into Bitcoin's spatiotemporal scaling laws, shedding light on the spending patterns of its users. Our findings underscore the potential value of anonymized, aggregate statistics about users' spatiotemporal spending behaviours, a prospect that could significantly benefit custodial exchanges, wallet software providers, and other cryptocurrency businesses. However, our work also serves as a stark reminder of the pressing need for robust network anonymity safeguards within the peer-to-peer layer of cryptocurrencies. As the cryptocurrency landscape evolves, the risk of the lack of anonymity being exploited against users looms larger. For example, nation-states may seek to tax Bitcoin users based on their involvement in the Bitcoin network. On a more optimistic note, the Bitcoin and the broader cryptocurrency community could already use a rich line of literature to enhance network anonymity~\cite{bojja2017dandelion,fanti2018dandelion++}.

\ifanonymous
\else
\paragraph{Acknowledgements.}
The research was co-funded by the project Strengthening the EIT Digital Knowledge Innovation Community in Hungary (2021-1.2.1-EIT-KIC-2021-00006), implemented with the support provided by the Ministry of Innovation and Technology of Hungary from the National Research, Development and Innovation Fund, financed under the 2021-1.2.1-EIT-KIC funding scheme.
\fi

\bibliography{sample}

\begin{thebibliography}{10}

\bibitem{apostolaki2021perimeter}
Maria Apostolaki, Cedric Maire, and Laurent Vanbever.
\newblock Perimeter: A network-layer attack on the anonymity of
  cryptocurrencies.
\newblock In {\em Financial Cryptography and Data Security: 25th International
  Conference, FC 2021, Virtual Event, March 1--5, 2021, Revised Selected
  Papers, Part I 25}, pages 147--166. Springer, 2021.

\bibitem{biryukov2014deanonymisation}
Alex Biryukov, Dmitry Khovratovich, and Ivan Pustogarov.
\newblock Deanonymisation of clients in bitcoin p2p network.
\newblock In {\em Proceedings of the 2014 ACM SIGSAC conference on computer and
  communications security}, pages 15--29, 2014.

\bibitem{bojja2017dandelion}
Shaileshh Bojja~Venkatakrishnan, Giulia Fanti, and Pramod Viswanath.
\newblock Dandelion: Redesigning the bitcoin network for anonymity.
\newblock {\em Proceedings of the ACM on Measurement and Analysis of Computing
  Systems}, 1(1):1--34, 2017.

\bibitem{brockmann2006scaling}
Dirk Brockmann, Lars Hufnagel, and Theo Geisel.
\newblock The scaling laws of human travel.
\newblock {\em Nature}, 439(7075):462--465, 2006.

\bibitem{de2015unique}
Yves-Alexandre De~Montjoye, Laura Radaelli, Vivek~Kumar Singh, and
  Alex~“Sandy” Pentland.
\newblock Unique in the shopping mall: On the reidentifiability of credit card
  metadata.
\newblock {\em Science}, 347(6221):536--539, 2015.

\bibitem{fanti2018dandelion++}
Giulia Fanti, Shaileshh~Bojja Venkatakrishnan, Surya Bakshi, Bradley Denby,
  Shruti Bhargava, Andrew Miller, and Pramod Viswanath.
\newblock Dandelion++ lightweight cryptocurrency networking with formal
  anonymity guarantees.
\newblock {\em Proceedings of the ACM on Measurement and Analysis of Computing
  Systems}, 2(2):1--35, 2018.

\bibitem{gao2021practical}
Yue Gao, Jinqiao Shi, Xuebin Wang, Ruisheng Shi, Zelin Yin, and Yanyan Yang.
\newblock Practical deanonymization attack in ethereum based on p2p network
  analysis.
\newblock In {\em 2021 IEEE Intl Conf on Parallel \& Distributed Processing
  with Applications, Big Data \& Cloud Computing, Sustainable Computing \&
  Communications, Social Computing \& Networking
  (ISPA/BDCloud/SocialCom/SustainCom)}, pages 1402--1409. IEEE, 2021.

\bibitem{juhasz2018bayesian}
P{\'e}ter~L Juh{\'a}sz, J{\'o}zsef St{\'e}ger, D{\'a}niel Kondor, and G{\'a}bor
  Vattay.
\newblock A bayesian approach to identify bitcoin users.
\newblock {\em PloS one}, 13(12):e0207000, 2018.

\bibitem{kapur13china}
Saranya Kapur.
\newblock China's google is now accepting bitcoin.
\newblock
  \url{https://www.businessinsider.com/chinas-google-is-now-accepting-bitcoin-2013-10},
  October 2013.
\newblock [Online; accessed 7-Sept-2023].

\bibitem{kelion13bitcoin}
Leo Kelion.
\newblock Bitcoin sinks after china restricts yuan exchanges.
\newblock \url{https://www.bbc.com/news/technology-25428866}, December 2013.
\newblock [Online; accessed 7-Sept-2023].

\bibitem{kondor2014rich}
D{\'a}niel Kondor, M{\'a}rton P{\'o}sfai, Istv{\'a}n Csabai, and G{\'a}bor
  Vattay.
\newblock Do the rich get richer? an empirical analysis of the bitcoin
  transaction network.
\newblock {\em PloS one}, 9(2):e86197, 2014.

\bibitem{koshy2014analysis}
Philip Koshy, Diana Koshy, and Patrick McDaniel.
\newblock An analysis of anonymity in bitcoin using p2p network traffic.
\newblock In {\em Financial Cryptography and Data Security: 18th International
  Conference, FC 2014, Christ Church, Barbados, March 3-7, 2014, Revised
  Selected Papers 18}, pages 469--485. Springer, 2014.

\bibitem{lischke2016analyzing}
Matthias Lischke and Benjamin Fabian.
\newblock Analyzing the bitcoin network: The first four years.
\newblock {\em Future internet}, 8(1):7, 2016.

\bibitem{nakamoto2008bitcoin}
Satoshi Nakamoto.
\newblock Bitcoin: A peer-to-peer electronic cash system.
\newblock {\em Decentralized business review}, 2008.

\bibitem{smirnov1948table}
Nickolay Smirnov.
\newblock Table for estimating the goodness of fit of empirical distributions.
\newblock {\em The annals of mathematical statistics}, 19(2):279--281, 1948.

\end{thebibliography}
\bibliographystyle{plain}
\end{document}